\documentclass[cits]{PoS}

\usepackage{xspace}

\newcommand{\eprest}{$E_{\rm p,rest}$\xspace}
\newcommand{\ep}{$E_{\rm p}$\xspace}
\newcommand{\eiso}{$E_{\rm{iso}}$\xspace}
\newcommand{\tninety}{$T_{90}$\xspace}
\newcommand{\tninetyrest}{$T_{90, \rm{rest}}$\xspace}

\newcommand{\batse}{{BATSE}\xspace}

\title{Rest-frame properties of gamma-ray bursts observed by the \emph{Fermi} Gamma-Ray Burst Monitor}

\ShortTitle{Rest-frame properties of GBM-GRBs}

\author{\speaker{David Gruber} on behalf of the \emph{Fermi}/GBM collaboration\\
          Max Planck Institute for extraterrestrial Physics, Giessenbachstr. 1., 
  \\85748 Garching, Germany\\
        E-mail: \email{dgruber@mpe.mpg.de}}

\abstract{In this talk I present the main spectral and temporal properties of \emph{Fermi}/GBM gamma-ray bursts (GRBs) with known redshift. Key properties of these GRBs in the rest-frame of the progenitor are investigated to better understand the intrinsic nature of these events.
The sample comprises 47 GRBs with measured redshift that were observed by GBM until May 2012. 39 sources belong to the long-duration population and 8 events were classified as short bursts. For all of these events we derive, where possible, the intrinsic peak energy in the $\nu F_{\nu}$ spectrum (\eprest), the duration in the rest-frame, defined as the time in which 90\% of the burst counts were observed (\tninetyrest) and the isotropic equivalent bolometric energy  (\eiso).
We confirm the tight correlation between \eprest and \eiso (Amati relation) with a larger scatter than previously reported. We also confirm the relation between \eprest and the 1-s peak luminosity ($L_p$) (Yonetoku relation). Short GRB 080905A, whose host galaxy was identified at redshift $z=0.1218$ is a peculiar outlier of this relation. Moreover, an intriguing, but preliminary, cosmic evolution of \eprest was observed, while no such evolution is evident for \tninetyrest. 
}

\FullConference{Gamma-Ray Bursts 2012 Conference -GRB2012,\\
		May 07-11, 2012\\
		Munich, Germany}

\begin{document}

\section{Introduction}

Gamma-ray Bursts (GRB), the most luminous flashes of $\gamma$-rays, are believed to originate from a compact source with highly relativistic collimated outflows ($\Gamma > 100$). A large fraction of our knowledge of the prompt emission comes from the Burst and Transient Source Experiment \cite{meegan92} onboard the Compton Gamma-Ray Observatory (\emph{CGRO}, 1991-2000). Unfortunately, only a handful of \batse bursts had a measured redshift. The lack of distance measurements led to a focus of GRB studies in the observer frame without redshift corrections. Due to the cosmological origin of GRBs, such a correction is likely to be necessary to understand the intrinsic nature of these events.

With the two dedicated satellites, Beppo-\emph{SAX} \cite{boella97} and \emph{Swift} \cite{gehrels04}, the situation has changed and afterglow and host galaxy spectroscopy has provided redshifts for more than 250 events by now. Unfortunately, the relatively narrow energy band of Beppo-\emph{SAX} (0.1~keV - 300~keV) and \emph{Swift}/BAT (15~keV - 150~keV) limits the constraints on the prompt emission spectrum \cite{butler07,saka09}. 
The \emph{Fermi} Gamma-Ray Burst Monitor (GBM) \cite{meegan09}, specifically designed for GRB studies, observes the whole unocculted sky 12 NaI scintillation detectors (8 keV to 1~MeV) and two BGO detectors (200~keV to 40~MeV).

Taking advantage of the broad energy coverage of GBM, the primary spectral and temporal properties, and energetics in the rest-frame of the progenitors of 47 GRBs with measured redshift are studied.

\section{Data analysis}

\begin{figure}[htbp]
\begin{center}
\includegraphics[width=0.6\textwidth]{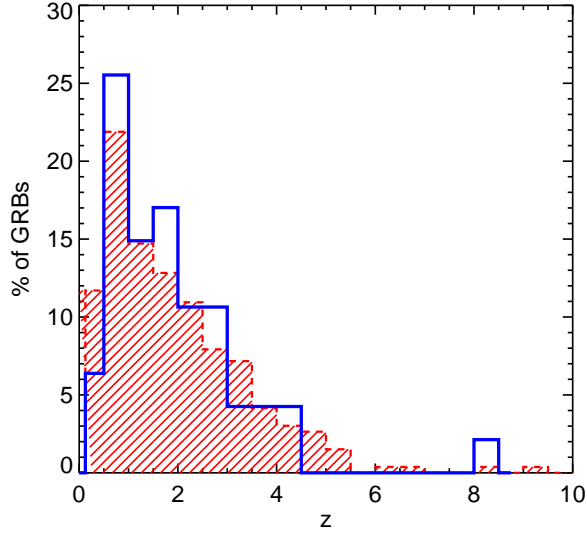}
\caption{Redshift distribution in \% of GBM GRBs (blue solid line) compared to all 256 GRBs with measured redshift to date (red dashed line). Both samples contain long and short bursts.}
\label{fig:zhist}
\end{center}
\end{figure}

The selection criterion for our sample is solely based on the
redshift determination. We form a sample of 47 bursts with known redshiftt\footnote{\tt{www.mpe.mpg.de/\textasciitilde jcg/grbgen.html}} detected by GBM up to May, 2012 (see Fig.\ref{fig:zhist}).

Four model fits were applied to all GRBs: a single power-law (PL), a power law function with an
exponential high energy cutoff (COMP), the Band function \cite{band93a} and a smoothly broken power law (SBPL). All models, except for the PL model, return a peak energy \ep. The data analysis was carried out according to, and consistent with, the GBM spectral catalogue \cite{goldstein12}. We were able to recover the \ep for 40 GRBs (34 long and 6 short GRB) of our sample (7 GRBs were best fit by a PL).

For determining the duration of a GRB, we applied the definition first introduced by \cite{kouv93}, i.e. the time in which 90\% of the burst counts is collected (\tninety). 
We determine the burst duration in count space in the rest-frame energy interval from  $100$~keV to $500$~keV, i.e. in the observer frame energy interval from $100/(1+z)$~keV to $500/(1+z)$~keV, also correcting for the time dilation due to cosmic expansion.
The cut between short and long GRBs was set artificially at the rest-frame duration of $2$~s, resulting in 8 short and 39 long GRBs.

\section{Correlations}
\subsection{Amati relation}
\label{subsec:amati}
\begin{figure}[htbp]
\begin{center}
\includegraphics[width=0.47\textwidth]{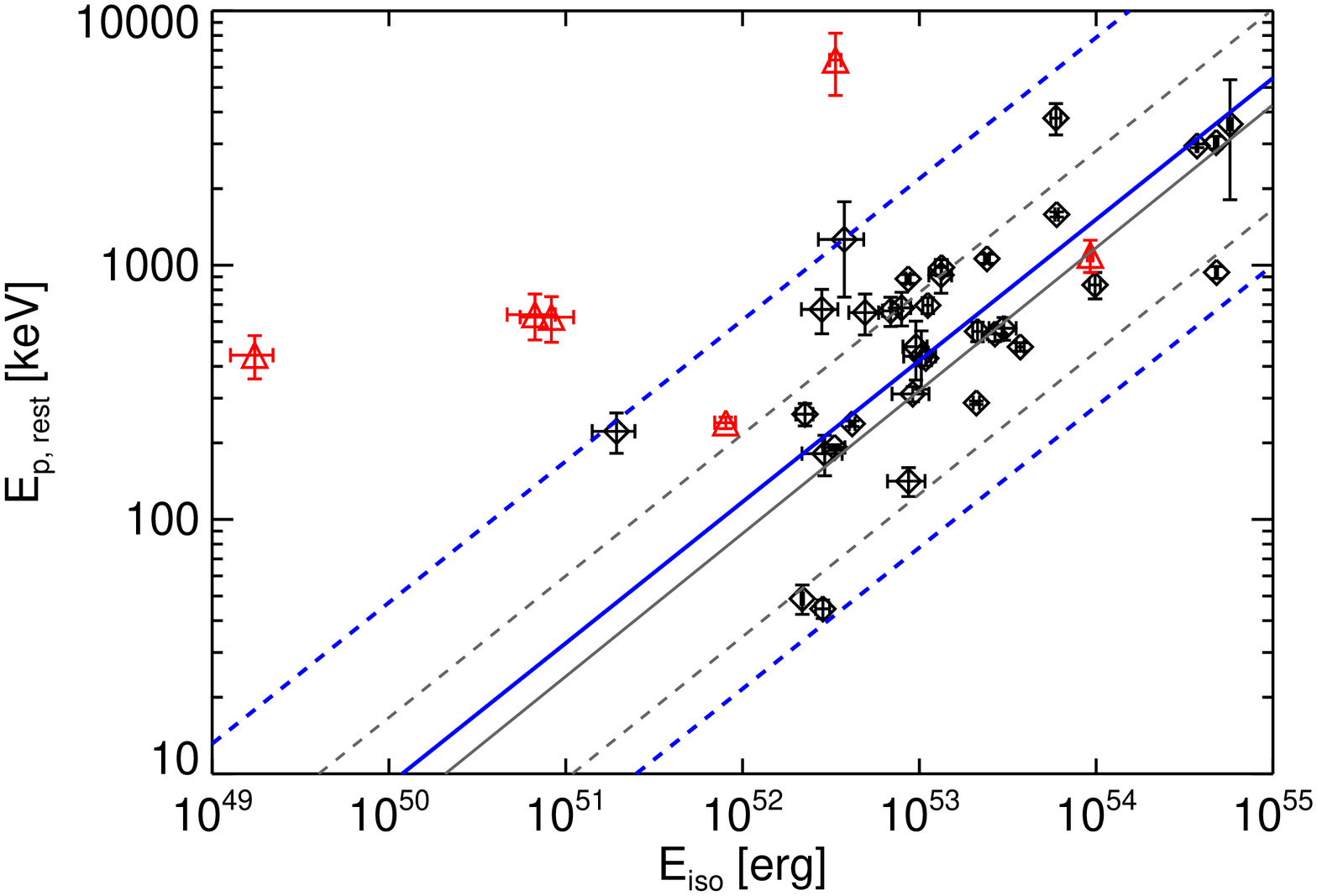} \hfill
\includegraphics[width=0.47\textwidth]{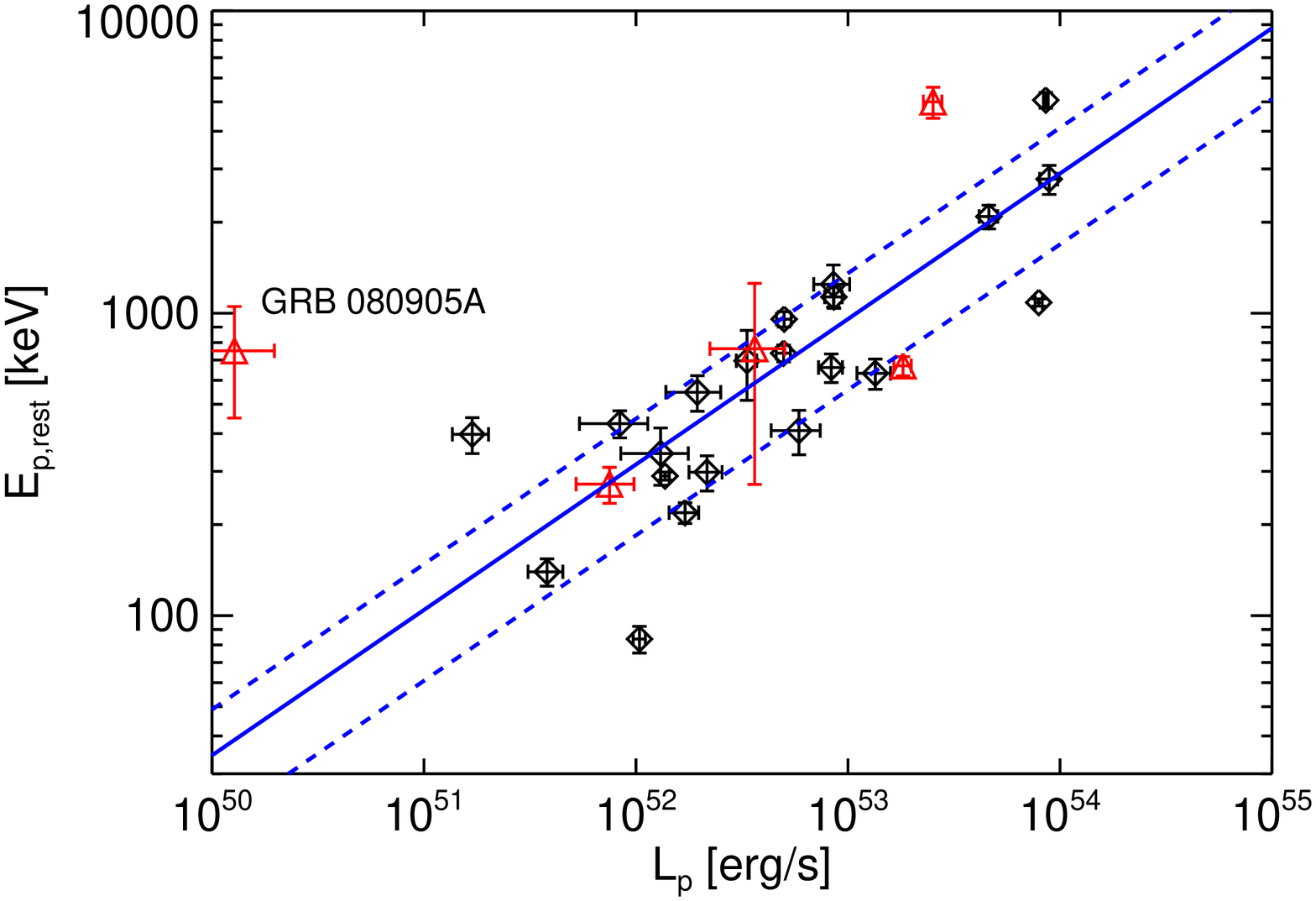} \\\
\parbox[t]{0.47\textwidth}{\caption{Amati relation for 6 short (red open triangles) and 34 long (black open diamonds) GBM GRBs. Also shown is the best power-law fit to the data (blue solid line) and the extrinsic $2\sigma$ scatter (blue dashed lines) with the best power-law fit published by \cite{amati10} (light-grey solid line) with the $2\sigma$ scatter (light-grey dashed line).}\label{fig:amati}} \hfill
\parbox[t]{0.47\textwidth}{\caption{Yonetoku relation for 5 short (red open triangles) and 21 long (black open diamonds) GBM GRBs. Also shown is the best power-law fit to the data (blue solid line) together with the $1\sigma$ scatter (blue dashed line).}\label{fig:yonetoku}}
\end{center}
\vspace{-0.7cm}
\end{figure}

It was shown by \cite{amati02} that there is a tight correlation between \eprest and \eiso. This ``Amati relation'' is shown in Fig.\ref{fig:amati} for  40 GBM GRBs with measured \eprest and \eiso. (\eiso was determined in the rest-frame energy range from 1~keV to 10~MeV, using the following cosmological parameters:  $\Omega_m=0.27$, $\Omega_{\Lambda}=0.73$ and $H_0=70.4$~km~s$^{-1}$~Mpc$^{-1}$). There is an evident correlation between these two quantities for long GRBs (Spearman's rank correlation of $\rho= 0.67$ with a chance probability of $1.73\times10^{-5}$).
Using the bisector of an ordinary least-squares fit (bOLS), we find 

\begin{equation}
E_{\rm p,rest} = 441^{+1840}_{-360} \times \left(    \frac{  E_{\rm{iso}}    }{     1.07\times10^{53} \,{\rm erg}    }  \right)^{0.55\pm0.10}\; \rm{keV}
\end{equation}

 which is in agreement with the indices obtained by e.g. \cite{amati10, ghina09, ghirlanda10} (errors refer to the $95$~\% CL). As has been shown by other authors in the past (see e.g. \cite{amati10, ghina09, amati08}) short bursts do not follow the relation, being situated well outside the 2~$\sigma$ scatter around the best-fit. This is true also for the power-law fit derived here (see Fig.\ref{fig:amati}) except for GRB~100816A and GRB~110731A. However, the former burst may actually fall in an intermediate or hybrid class of short GRBs with extended emission (see e.g. \cite{norris06, zhang09}) while the latter is short only in the rest-frame.

\subsection{Yonetoku relation}
A tight correlation between \eprest  and the 1-s peak luminosity $(L_p)$ in GRBs was found by \cite{yonetoku04} (so called Yonetoku relation). We determined $L_p$ and the time resolved \eprest in the brightest $1024$~ms and $0.064$~ms time bin for long and short GRBs, respectively.
We were able to determine the time resolved \eprest for 26 (5 short and 21 long) GRBs and we present this relation in Fig.~\ref{fig:yonetoku}. Using again a bOLS, omitting short GRB 080905A from the fit (see below), we find 

\begin{equation}
E_{\rm p,rest} = 667^{+295}_{-310} \times \left(    \frac{  {L_p}    }{     4.97\times10^{53}\,{\rm erg}\,{\rm s}^{-1}   }  \right)^{0.48\pm 0.01}\; \rm{keV},
\end{equation}

with the errors referring to the $68$~\% CL. 

The Spearman's rank correlation gives $\rho=0.8$ with a chance probability of $9\times10^{-8}$. Our findings are in good agreement with \cite{ghina09, ghirlanda10} and \cite{yonetoku04}.
We note that short GRB~080905A is a peculiar outlier to the Yonetoku relation. The redshift for this GRB was obtained via its claimed host galaxy at $z=0.1218$ \cite{rowlinson10}, making it the closest short GRB to date. It could be that this GRB has some peculiar properties compared to other GRBs in order to explain its position in the \eprest - $L_p$ diagram. Another possibility is that it may have actually occurred at a higher redshift and that the claimed host is only a foreground galaxy, even though \cite{rowlinson10} find a chance alignment probability of $< 1$~\%. (A redshift of $z\sim0.9$ would make GRB~080905A consistent with the Yonetoku relation.)

\section{Cosmic evolution}
\subsection{\tninetyrest vs redshift}
\begin{figure}[htbp]
\begin{center}
\includegraphics[width=0.47\textwidth]{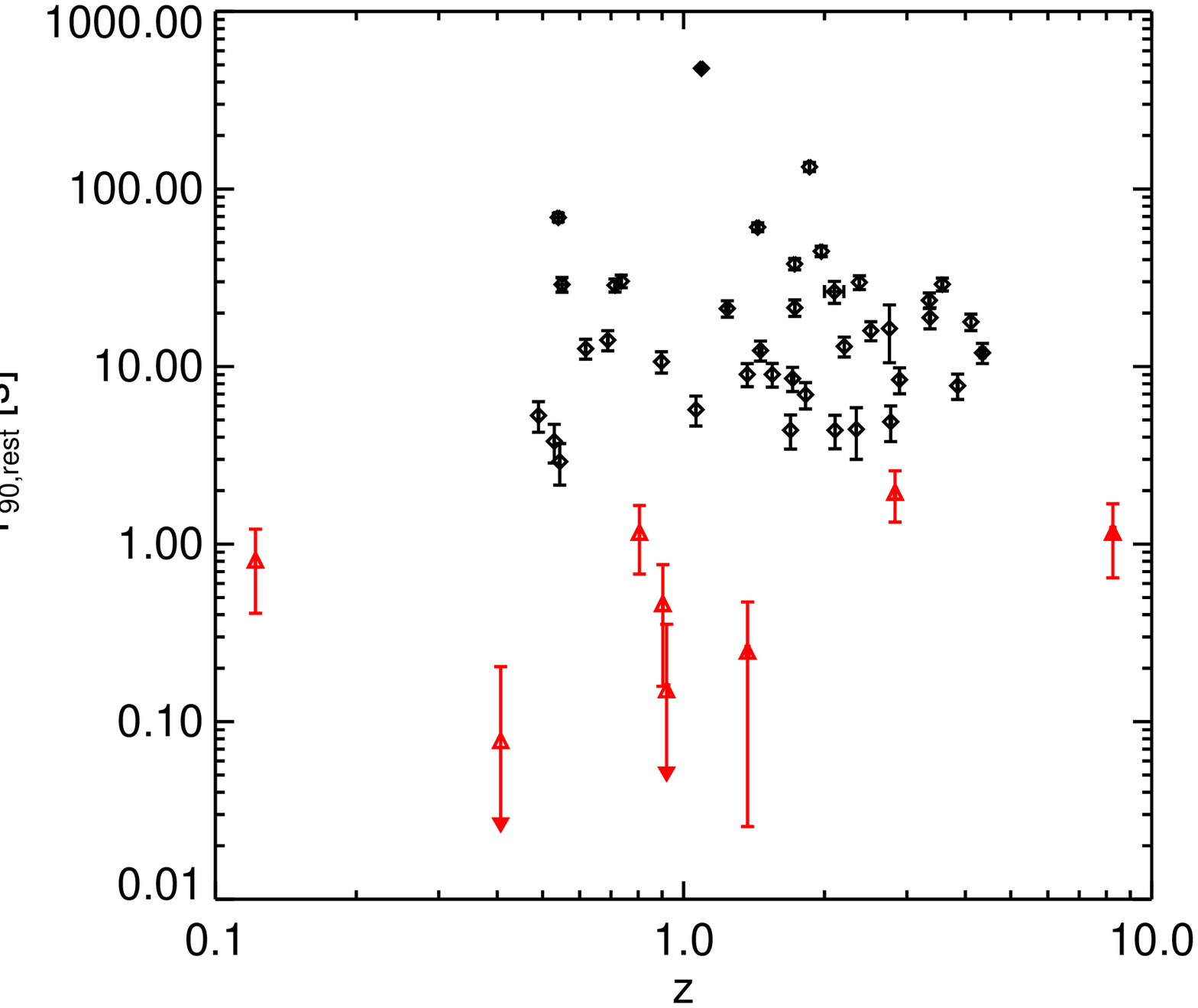}
\includegraphics[width=0.47\textwidth]{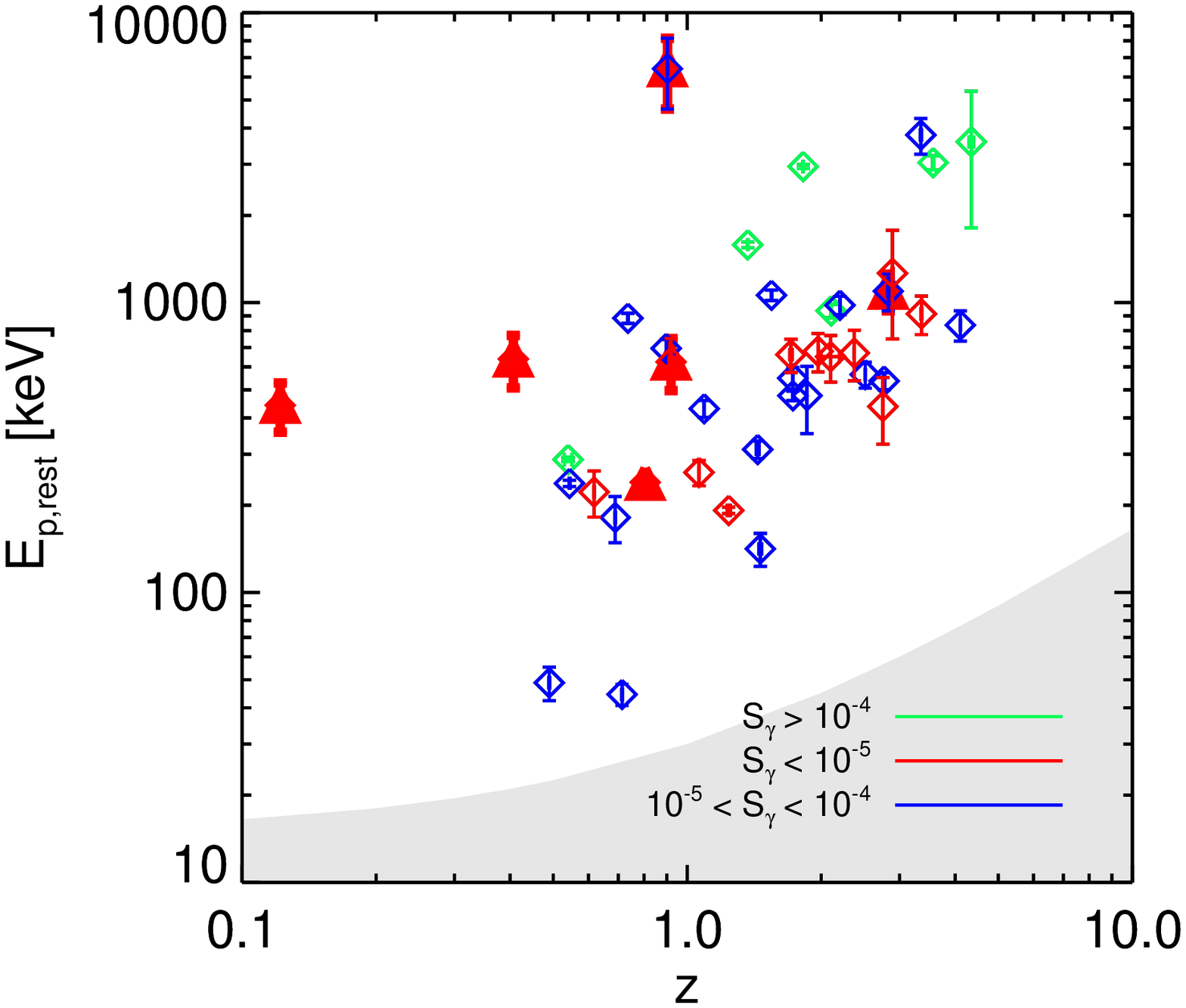}
\parbox[t]{0.47\textwidth}{\caption{Testing the cosmic evolution of \tninetyrest for long (black diamonds) and short (red triangles) GRBs.}\label{fig:t90vsz}} \hfill
\parbox[t]{0.47\textwidth}{\caption{Testing the cosmic evolution of \eprest for long (black diamonds) and short (red triangles) GRBs. The grey-shaded region indicates the area of \eprest which cannot be probed by GBM. Bursts with high (green), intermediate (blue) and low (red) fluence (in erg~cm$^{-2}$) are labeled.}\label{fig:yonetoku}}
\end{center}
\end{figure}

Several authors (e.g. \cite{kocevski11}) reported that, due to the detector sensitivity the observed duration can actually decrease with increasing redshift as only the brightest portion of a high redshift GRB's light curve become accessible to the detector. Consequently, this would mean that probably all estimates of duration and subsequently energetics for high redshift GRBs are only lower limits to their true intrinsic values. Indeed, \cite{pel08} find such a negative correlation between \tninetyrest and $z$. However, contrary to these authors, we do not find any evidence in the GBM data for any dependence of \tninetyrest on $z$ (see Fig.\ref{fig:t90vsz}). Our results confirm the analyses with \emph{Swift} detected GRBs \cite{greiner11}.

\subsection{\eprest vs redshift }

In order to explain the detection rate of GRBs at high-$z$, \cite{salva07} conclude that high-$z$ GRBs must be more common (e.g. \cite{daigne06,wang11}) and/or intrinsically more luminous \cite{salva09} than bursts at low-$z$ (but see \cite{but10}). Assuming that the luminosity function of GRBs indeed evolves with redshift and that the Yonetoku relation is valid, we would also expect a positive correlation of \eprest with $z$.

A Spearman's rank correlation test of \eprest and $z$, using only the long GRBs, results in $\rho=0.67$ with a chance probability of $P=1.3\times10^{-5}$. While such a significant correlation surely is intriguing, it was already argued by \cite{gruber11} that selection effects need to be taken into account before a reliable claim on redshift-evolution of \eprest can be made.

\section{Conclusion \& Summary}

Here, the data and analysis of 47 GRBs with redshift that were observed by \emph{Fermi}/GBM was presented. The main focus was laid upon the temporal and spectral properties, as well as on the energetics and intra-parameter relations within these quantities.
The \eprest - \eiso correlation was confirmed and a power law with index of $0.55$ was found to adequately fit the data. Although this is consistent with the values reported in the literature, the scatter around the best-fit is significantly larger. We also confirm a strong correlation between $(L_p)$ and its \eprest with a best-fit power law index of $0.48$.

There is no observed redshift evolution of \tninetyrest, whereas there might be some indication that \eprest of long GRBs is higher at higher redshifts. This result, however, is heavily influenced by selection effects which need to be taken properly into account before making any conclusive statement about this effect.

Finally, we report that short GRB~080905A is a striking outlier to the Yonetoku relation. Either because it is a GRB with peculiar properties compared to other long and short GRBs, or because the identified host galaxy is in fact a foreground object and not related to the burst emission site.

%\bibliographystyle{aa}
%\bibliography{references}

%Author(s), {\\emph Title}, {\\emph Journal} {\\bf Volume} (Year) Page 

\end{document}